# Integrated micro-comb sources for quantum optical applications

Michael Kues,[1,*] Christian Reimer,[2] Joseph M. Lukens,[3] William J. Munro,[4] Andrew M. Weiner,[5] David J. Moss,[6] and Roberto Morandotti[7,*]

[1] School of Engineering, University of Glasgow, Rankine Building, Oakfield Avenue, Glasgow G12 8LT, UK; Department of Engineering, Aarhus University, Finlandsgade 22, 8200 Aarhus N, Denmark

[2] John A. Paulson School of Engineering and Applied Sciences, Harvard University, Cambridge, 02138, USA

[3] Quantum Information Science Group, Computational Sciences and Engineering Division, Oak Ridge National Laboratory, Oak Ridge, Tennessee 37831, USA

[4] NTT Basic Research Laboratories and NTT Research Center for Theoretical Quantum Physics, NTT Corporation, Kanagawa, Japan; National Institute of Informatics, Tokyo, Japan

[5] School of Electrical and Computer Engineering and Purdue Quantum Center, Purdue University, West Lafayette, IN 47907, USA

[6] Centre for Micro Photonics, Swinburne University of Technology, Hawthorn, Victoria 3122, Australia

[7] Institut National de la Recherche Scientifique (INRS-EMT), 1650 Blvd. Lionel-Boulet, Varennes, J3X1S2, Canada; Institute of Fundamental and Frontier Sciences, University of Electronic Science and Technology of China, Chengdu 610054, China; ITMO University, St Petersburg, Russia

[*] Corresponding authors

**Abstract** (120 words)

A key challenge for quantum science and technology is to realise large-scale, precisely controllable, practical systems for non-classical secured communications, metrology and ultimately meaningful quantum simulation and computation. Optical frequency combs represent a powerful approach towards this, since they provide a very high number of temporal and frequency modes which can result in large-scale quantum systems. The generation and control of quantum optical frequency combs will enable a unique, practical and scalable framework for quantum signal and information processing. Here, we review recent progress on the realization of energy-time entangled optical frequency combs and discuss how photonic integration and the use of fiber-optic telecommunications components can enable quantum state control with new functionalities, yielding unprecedented capability.

**Introduction**

The generation of large-scale quantum systems that can be precisely controlled is critical[1–3] to enable practical and high-performing devices based on quantum technology that move beyond proof-of-concept demonstrations. These systems can be used for enhancing link speeds in non-classical communications[4], enabling practical quantum enhanced metrology[5], and for increasing the quantum information content in computations and simulations[6]. Starting from the typical two-dimensional 'qubit' (the quantum analog of a classical bit) a higher degree of complexity can be achieved through the use of multi-partite states (i.e. formed by multiple sub-systems, such as atoms, electrons or photons and their degrees of freedom) or by increasing the number of quantum dimensions of each party. The Hilbert space size (analog to the available quantum resource) scales as $d^N$ (with d and N being the number of dimensions and parties, respectively). The polynomial scaling with dimensionality results in a favourable increase in state complexity for smaller dimensions, but becomes less efficient for larger ones, compared to increasing the number of parties. Further, in addition to boosting the quantum information content, complex states may provide a reduced sensitivity to noise – an ever-present challenge for quantum systems – and can enable novel algorithms. For example, two-level 'cluster states' carry unique multi-partite entanglement properties that allow the implementation of universal quantum computation via the 'one-way' scheme[7,8], where processing is performed through measurements. Increasing dimensions via high-dimensional sub-systems – so-called 'qudit' states – can additionally yield higher noise robustness[9] and allow new algorithms for high-dimensional quantum computation[10].

Developing suitable non-classical systems is, however, pushing the limits of current technology. For example, atomic[11] and superconducting[12] approaches, traditionally based on qubit schemes, need ever more technologically complex systems and control methods to increase the number of qubits – a result of their intrinsic physics as well as interactions between neighbours and their environment. For these systems, scaling beyond qubits into qudits is difficult due to their intrinsically shorter decoherence times as well as limited gate fidelities for higher quantum levels.

Photonic systems,[13] on the other hand, are attractive not just because photons can interact with other quantum systems, can transmit quantum information over long distances and operate at room temperature, but also because photonics can support the generation of very complex quantum states and facilitate their precise control, including high-dimensional systems. Initial approaches based on encoding quantum information on spatial, polarization, and orbital angular momentum modes[8,14,15] have been very successful at creating high-dimensional optical states[15,16] as well as qubit cluster-states[8,17]. However, despite their success, the footprint of these approaches necessarily grows as the dimensionality increases, due to the spatially distinct nature of the exploited degrees of freedom. This ultimately poses a significant limitation to the degree of complexity that is achievable for practical systems.

The recent introduction of 'quantum frequency combs' can enable the scalable generation of complex (i.e. multi-partite and high-dimensional) states in the frequency domain[18–22]. Their multi-modal characteristics – having many phase stable frequency modes residing in a single spatial mode – allows these frequency structures to store quantum information in the joint bi-photon complex spectral amplitude and phase, as well as in temporal and frequency correlations. Consequently, quantum frequency combs can store and process a large amount of information in their spectro-temporal quantum modes, quickly becoming a rapidly growing field of research[18–32].

The first quantum frequency combs were based on free-space optical parametric oscillators (OPOs) using nonlinear crystals, where phase-matched parametric down conversion leads to quantum correlations between frequency modes and results in the generation of so-called 'squeezed states' in a continuous variable (CV) formalism[33,34]. This approach[18,23,24,29–32,35] generates very large-scale photonic quantum systems with mode counts exceeding one million. In synchronously excited systems, this can lead to multi-mode entangled networks[30,32] which can generate highly complex cluster states[23–26,35,36]. These, however, have proven difficult to control since their very narrow free spectral range (frequency mode spacing) of typically only ~ 100 MHz makes accessing individual modes very challenging, although some reports have achieved state characterization using homodyne detection for quantum combs with up to 1 GHz spacing[29]. Consequently, multiple modes are often grouped and manipulated together. Moreover, their large and complex setups require sophisticated stabilization schemes, and so these systems are not yet fully suitable for applications outside of the lab. An important distinction between photon and squeezed states is that the latter are Gaussian quantum states[37], and therefore cannot be used for nontrivial quantum information processing in combination with exclusively Gaussian operations and measurements[38–40], such as homodyne detection. In order to use squeezed states for e.g. quantum computation, non-Gaussian operations and measurements will have to be implemented, such as photon-number resolving detection. In contrast, entangled photon states combined with coincidence detection can be directly used for quantum information processing.

Integrated photonics is a powerful approach that can allow fundamental advances in scalable, classical[41] and quantum[42,43] photonic systems beyond what is achievable using bulk optics. Further, integrated chips that are compatible with the silicon chip industry offer the greatest advantages in terms of being mass-producible, stable, low-cost and practical[41,44]. Integrated micro-combs, or Kerr combs, in particular, have found substantial interest since their first report[45–48] about 10 years ago, enabling the miniaturization of classical frequency combs[49,50]. Their quantum counterpart – 'quantum micro-combs' – operating at the single-photon level, has attracted significant attention in addressing many of the challenges outlined above[19,20,51].

With their unique frequency signature, quantum-micro combs can enable photon entanglement in discrete frequency modes that allows for the generation and control of quantum states with considerably enhanced complexity. They achieve this in a scalable, manufacturable platform for generating frequency-multiplexed heralded photons and multi-photon, high-dimensional and hyper-entangled states. This quantum scalability comes from their ability to operate in the frequency picture where more frequency modes and higher dimensions can be easily added to the system without significant overhead or penalty. In contrast to continuous-variable systems, working with discrete-variable photon states is compatible with post-selection using advanced single-photon detection technologies, where losses do not intrinsically degrade the measured quantum state; loss does, however, result in the practical reduction of detection rates. Importantly, quantum photonic systems operating near 1550 nm also tremendously benefit from, and can build on, well-established telecommunications technology, allowing the realization of chip/fiber-based quantum devices and distributed fiber quantum networks. Frequency-encoded complex quantum states of light can be readily and directly manipulated using state-of-the-art optical fiber components[20,22,52,53] including filters, routers, and modulators. Taken together, all of these factors make quantum micro-combs a compelling approach to enabling compact, controllable, and scalable complex quantum systems.

Here, we review quantum frequency combs that operate via photon entanglement, beginning with mode-locked quantum frequency combs,[28] followed by energy-time entanglement methods including their generation and control. We focus on integrated optical solutions that operate via the generation of photons using nonlinear optical processes directly in micro-cavities, as well as via passive spectral filtering following the generation of photons in nonlinear optical waveguides. Finally, we discuss their potential for future applications, addressing the challenges for photonic-based quantum science.

**Frequency combs occupied by few photons**

Classical frequency combs are characterized by coherently oscillating light fields over many spectral lines, each typically occupied by millions of photons or more. Their power lies in the phase coherence among the spectral lines achieved via mode-locking approaches, which leads to ultra-high stability – the basis for applications in metrology, spectroscopy and more. If a classical frequency comb (i.e. pulse train, or any classical coherent light field) is attenuated to the single photon level, the individual photons keep their spectral phase and temporal pulse shape, which enables, amongst other applications, their use for quantum communications[54]. More importantly, in addition to this single-photon spectral phase, frequency combs formed by few photons can carry two- or multi-photon spectral phase coherence, which is fundamentally different from single-photon spectral phase coherence. Bi-photon phase coherence can occur when two photons are generated from the same process, such as spontaneous parametric down conversion (SPDC, a second-order nonlinear effect), or spontaneous four wave-mixing (SFWM, a third-order effect). In SPDC and SFWM a nonlinear light-matter interaction mediates the annihilation of one or two photons from an excitation field, simultaneously generating two daughter photons, typically referred to as 'signal' and 'idler' photons. Photons generated via spontaneous parametric nonlinear processes, either directly within a resonator or in a waveguide (and subsequently subjected to periodic spectral filtering), exhibit discrete spectral modes and form a quantum frequency comb (Fig. 1).

The difference between single versus bi-photon phase coherence can be understood by considering an analogy to the well-understood single- and bi-photon polarization signatures. If a classical linearly polarized light field is attenuated to the single photon level, the individual photons remain linearly polarized, but if a polarization entangled photon pair is generated, while the single-photon polarization of each individual photon is random, the bi-photon polarization of the state is perfectly defined[55]. In analogy to this, it is possible to generate an energy-time entangled bi-photon state where the spectral phases of the individual photons in the state are random, but the two-photon spectral phase is perfectly defined and thus is stable (also referred to as a 'mode-locked two-photon state'[28]).

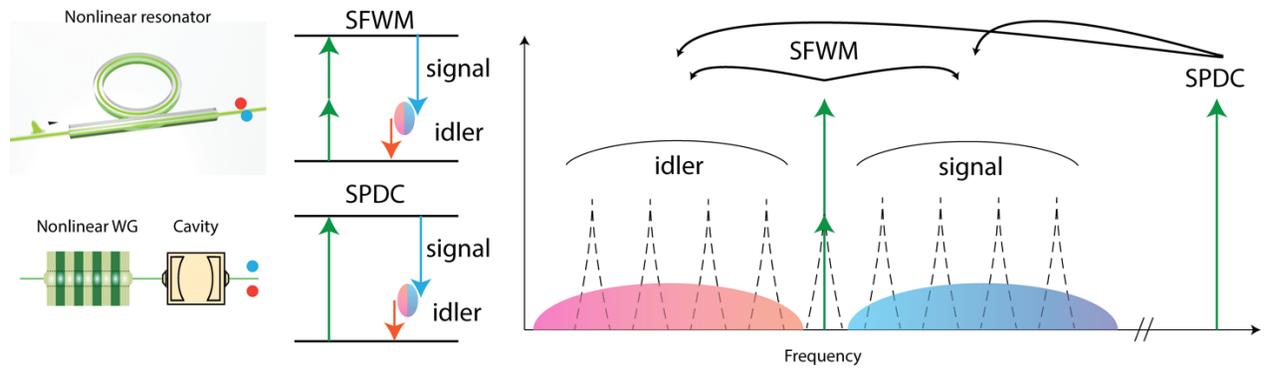

**Figure 1 |** Bi-photon emission of two correlated photons (called signal and idler) through spontaneous parametric down conversion (SPDC) or spontaneous four-wave mixing (SFWM) from filtered nonlinear waveguides or nonlinear resonators. Such process leads to discrete energy-time correlated bi-photon frequency combs.

We first consider the fundamental process that governs the generation of photons via nonlinear optical interactions in non-resonant structures such as waveguides and nanowires. The nonlinear parametric interaction that gives rise to photon pairs preserves the energy, momentum and polarization of the incident optical field, creating two photons in a continuum of temporal and frequency modes. If the phase-matching bandwidth for the process is larger than the excitation field's bandwidth, the joint spectrum and temporal signature of the two photons are correlated. This can be intuitively understood by noting that the spontaneous (i.e. non-deterministic) process is effectively simultaneous[56] within the bandwidth limitation of the process (determined by phase-matching) which, together with energy-conservation, directly leads to time-frequency Einstein-Podolsky-Rosen (EPR) correlations, enabling one to witness energy-time entanglement directly using uncertainty relations.[57] Energy-time correlated photon pairs have been generated in many ways, including in non-resonant waveguides using quasi-phase-matched periodically poled lithium niobate (PPLN)[57] or silicon waveguides designed, through their geometry, to provide phase-matching for SFWM[58], as well as in a range of other integrated devices[43,59–64]. In these cases, the simultaneous generation of two photons immediately dictates that if one photon is detected, the other photon has to arrive at the same time (within the temporal duration of the photon), leading to a strong correlation in the two-photon joint temporal intensity (JTI) (Fig. 2a). In addition, this process preserves the energy of the excitation photons. Since the photon energy is directly related to the frequency, the sum of the signal and idler frequencies is constant and so energy conservation yields an anti-correlation in the joint spectral intensity (JSI) (Fig 2b). Indeed, the full quantum state can be described by its joint spectral/temporal amplitude, which also includes phase terms. In the coincidence measurements shown in Fig. 2, only the intensities were measured, while phase coherence could be probed using alternative interferometric techniques, or broadband homodyne detection. In addition to individual measurements of the JTI and JSI[65], it is possible to simultaneously exploit the correlation in time and anti-correlation in frequency, leading to quantum interference in a Franson-type interferometer[66], for example. In particular, if the photons pass through a set of unbalanced interferometers, quantum interference in their coincidences arises that can prove energy-time entanglement of the photon state[66]. Optical interferometers are therefore a common tool to characterize energy-time entangled states. Energy-time entanglement has been demonstrated in systems using a CW excited silicon ring resonator where only two single resonator modes were considered[59]. It is customary to characterize this entanglement with the 'Schmidt-number' which is the number of significant orthogonal modes derived from a singular value decomposition of the joint spectral amplitude, and which gives a measure of how many frequency-time modes are entangled[67,68].

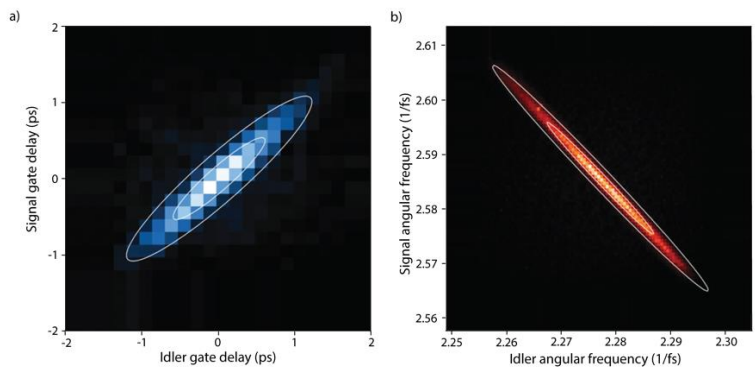

**Figure 2 | Correlation signature of time-energy entangled photons.** Joint a) temporal and b) spectral intensities of an energy-time entangled photon pair (adapted from Ref. [65]).

To achieve discretization in the spectrum, continuous energy-time entangled photons can be generated in a waveguide and then filtered by a cavity, yielding a frequency comb structure. Conversely, rather than filtering the photons after their generation, which leads to high losses in photon flux, it is possible to directly generate photons within a nonlinear resonator[19,28] to avoid filtering losses and at the same time exploit cavity enhanced parametric processes.[69] This can increase the pair generation efficiency, which is particularly useful for integrated devices since they mainly operate via the less efficient SFWM process. In order to enable spectral access to the individual modes, typically achieved via optical filtering, a large mode spacing is advantageous. The use of micro-resonators as a nonlinear medium is ideal for this, and can been accomplished in a variety of ways, including whispering-gallery mode resonators in $LiNbO_3$[70] and integrated ring resonators in a range of platforms such as SOI,[59,71] Hydex,[19,72,73] AlN,[63] and silicon nitride.[20,51] If the SFWM phase-matching is designed so that the bandwidth covers several resonances, which occurs with low anomalous second-order dispersion, it is possible to generate photon pairs over multiple frequency modes.[19–21] Figure 3a shows a typical photon comb spectrum spanning the entire telecommunications band with a 200 GHz frequency spacing, emitted from a silica-doped micro-ring resonator[21]. This large frequency spacing – only achievable with micro-cavities – allows the use of standard telecom filters (e.g. dense wavelength division multiplexing DWDM filters) to select the resonance modes individually for coincidence measurements, for example. Figure 3b depicts a typical correlation matrix resulting from a JSI measurement of the source, showing that coincidences are only detected for frequency components that are symmetrically located with respect to the excitation field[19,20,22,51]. This reflects the energy conservation of the process and thus the quantum frequency comb nature of the generated states. These systems can be directly used as frequency multiplexed heralded single photon sources (where the idler photon detection indicates the presence of the signal photon)[19], with applications to quantum key distribution[74], for example, or in combination with frequency shifting schemes for the generation of more deterministic single photon sources[75,76]. Furthermore, the modes of quantum frequency combs can be utilized as independent sources for multiplexing with WDM schemes for entanglement-based quantum communication systems[71].

Looking more closely at the entanglement signature of these emissions, the spectral correlation is no longer continuous but rather exhibits discrete modes (Fig. 3b). The photon lifetime in this case is no longer determined by the bandwidth of the nonlinear process, but by the lifetime of the resonator mode[70,77]. It has been shown that for a cavity filtered bi-photon spectrum this type of comb structure also manifests itself in quantum interference[27], observed in a Franson interferometer, for example. This leads to the disappearance and reappearance of quantum interference for interferometric imbalances corresponding to multiples of the inverse comb spacing[27,78]. While it is not unexpected that energy-time entanglement between different filtered spectral modes will occur when they are generated by passive filtering of a continuous spectrum, it has not been clear what role the vacuum fluctuations play in photon comb generation when it directly takes place in a nonlinear resonator, and whether energy-time entanglement between spectral modes would still arise in this case. Recent experiments show that energy-time entangled states directly generated in a resonator structure intrinsically cover all frequency modes, which was experimentally confirmed, in a similar fashion, by using optical interferometers with a variable delay[51]. However, it is still under debate whether interferometer-based time domain measurements showing the revival of interference imply that the bi-photon phase is stable[27,51].

A deeper understanding of bi-photon comb entanglement needs to consider the timescales of the processes relative to both the resonator bandwidths and nonlinear effects taking place. Generally, the extent of entanglement between the discrete resonator modes is determined by the over-all bandwidth of the nonlinear process and the frequency discretization is often referred to as frequency-bin entanglement[79]. If the excitation field bandwidth is smaller than the resonance bandwidth, the photon emission from a single resonance is multi-mode in frequency, in which case continuous energy-time entanglement additionally occurs for a single resonance pair. In this case the Schmidt mode number, which can be quantified through classical seeding experiments or single photon autocorrelation measurements, is much larger than unity[77,80,81]. On the other hand, when the bandwidth of the excitation field is similar to, or larger than, the resonance bandwidth, as is the case for a short-pulsed excitation, energy-time entanglement within a single resonance almost vanishes, yielding a Schmidt mode number close to unity[77]. However, by exciting the system with double or even multiple pulses with delay large enough to display a comb structure within a single resonator linewidth, a discretization in the time domain can be achieved. This excitation can be used to generate photons that additionally have a time-bin entanglement signature[21]. Consequently, multiplexed systems for time-bin and continuous energy-time entanglement have been achieved[71,82].

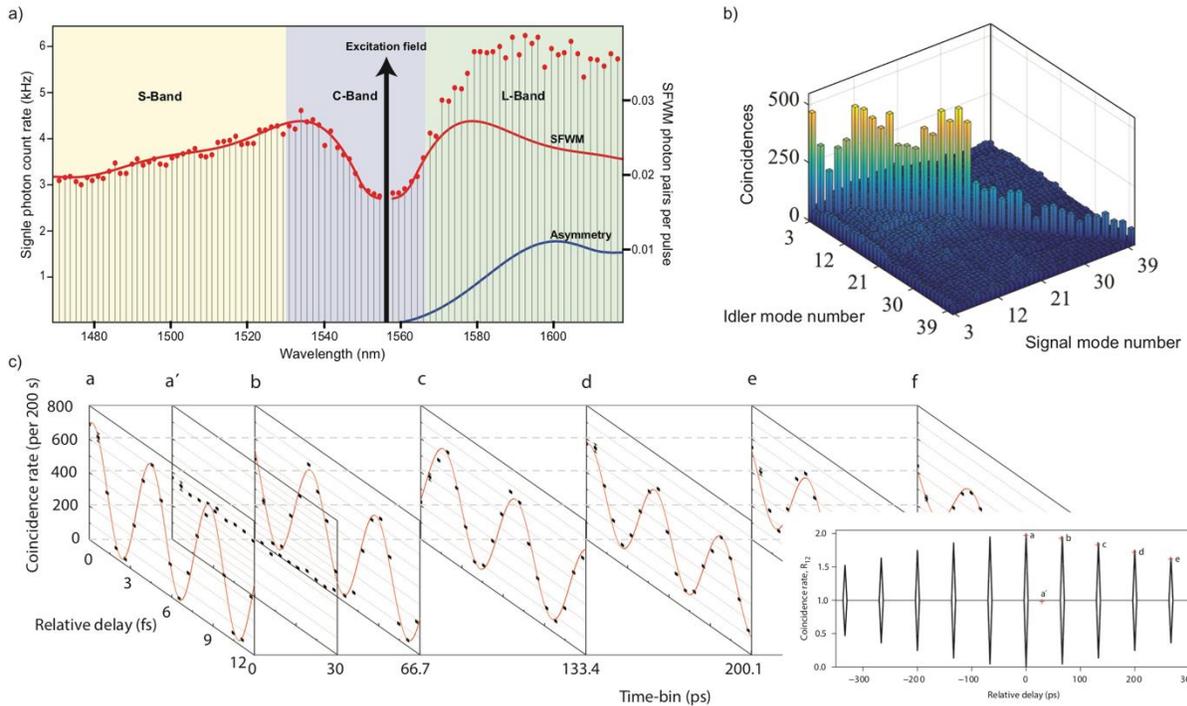
8

**Figure 3 | Bi-photon quantum frequency comb characteristics** a) Single photon spectrum emitted from a micro-ring resonator (adapted from Ref. 21). b) Discrete correlation matrix from a nonlinear ring resonator (adapted from Ref. 20). c) Quantum interference revivals for bi-photon combs demonstrating energy-time entanglement covering multiple resonances (adapted from Ref. [27]).

**Complex quantum states generated by micro-combs**

For many quantum information applications, it is required to have experimental access to all orthogonal modes of the system. Time- and frequency-bin systems are very advantageous in allowing this since their generation and control is readily supported by current laser and modulator technology. Time-bin systems benefit from pulsed lasers and interferometer technologies, whereas frequency-bin systems benefit from telecom filtering and radio frequency (RF) electro-optic spectral mixing. The practical and scalable implementation of time-bin and frequency-bin encoding, particularly when accomplished using micro-cavities, can lead to the generation of complex quantum states[20–22,52,53] by allowing an increase in the number of state parties (i.e. the number of independent sub-systems), through more photons or hyper-entangled approaches, and/or by allowing the practical increase of the state dimensionality.

Entangled multi-photon states can be achieved by time-bin encoding of resonator frequency combs. For this, two-photon time-bin entangled states are generated by exciting a nonlinear cavity, resonantly matched with two or more laser pulses. The excitation power is kept low enough so that each excitation pulse has a low probability ($< 0.1$) of generating a photon pair, meaning that the photon pair is created in a superposition of the time modes defined by the double pulse. This leads to a time-bin entangled two-dimensional state – i.e. an entangled two-qubit state. To characterize and verify the entanglement of this discrete state, projection measurements on the individual temporal modes, as well as their superpositions, are required. This can be achieved by sending signal and idler photons individually through an imbalanced interferometer, where both interferometers have similar arm length mismatches that coincide with the temporal separation of the pump pulses[21]. The indistinguishability between the creation time of the photon pair and which path of the interferometer (the long one or short one) the photons go through, gives rise to quantum interference, i.e. a modulation of the coincidence counts when changing the interferometer phase[21]. The interference visibility can be connected to a Bell inequality that, if violated, demonstrates EPR-like correlations and entanglement[66]. Due to phase-matching spanning several resonances of the cavity, time-bin entangled photon pairs can be generated over many resonance pairs simultaneously, leading to a comb of time-bin entangled two-photon states.

The distinctive multimode characteristic of the frequency comb also allows for the generation of four-photon time-bin entangled states. This can be achieved by exploiting two SFWM processes occurring within the coherence time of the pulsed excitation and then post-selecting two signal and idler pairs on different resonances simultaneously. The realization of this four-photon entangled state can be confirmed through quantum interference as well as quantum state tomography[21] (see Fig. 4). Furthermore, it should be possible (by selecting more resonances) to scale to even higher photon numbers of multi-photon states, thereby further increasing the state complexity and Hilbert space size. However, the drawback of multi-photon states is that their detection through post-selection is exponentially reduced by loss as the number of photons increases. A different, and complementary, way to increase the state Hilbert space size while maintaining low photon numbers is to use qudits.

Schemes that make use of the coherence between frequency modes can provide access to high-dimensional entangled states and enable encoding of quantum information on discrete frequency bin states. Here, the photon pairs are explicitly considered to be in a quantum superposition of many frequency modes[20,22,53]. The experiments that employed interferometers, as discussed above, are not capable of performing projections onto photon states with defined spectral phases within the comb. Furthermore, while broadband single photon detectors are able to detect photons over a wide frequency band, they are too slow to resolve ultra-fast temporal dynamics (on the order of a few picoseconds or less) that result from the phase coherence between different spectral modes. Ultrafast sum frequency generation, essentially the quantum analogue of intensity autocorrelation methods used extensively in classical ultrafast optics, has been used to resolve the time correlation function of broadband entangled photons with femtosecond resolution[83–88]. However, this approach is limited to local detection (both entangled photons must interact in the same nonlinear crystal), and signal levels are usually too low for use with quantum frequency combs. In the context of squeezed states, broadband detection techniques based on parametric amplification and a nonlinear interferometer configuration have been developed[89,90] to measure the complete joint spectral amplitude in the frequency domain simultaneously across an ultra-wide spectrum. However, the feasibility of transferring such techniques to photon states has not been studied yet. To do this, methods to access superpositions of frequency components would be needed.

In analogy to a polarization entangled state, where a single photon can be projected onto specific polarization states by means of phase plates and polarizers, the individual photons of a frequency-bin entangled state need to be projected onto specific spectral phase states. This can be achieved through the use of telecommunications components, many available off-the-shelf such as electro-optic phase modulators (see e.g. Fig. 4c). When driven with a sinusoidal RF signal these devices can generate sidebands that can be used to superimpose different frequency components of the generated photons. Individual manipulation of the frequency component's phase is possible via programmable optical filters based on optical pulse shaping technology[91], also commonly used in telecom networks. Combining these elements allows one to perform single photon projections on any arbitrary spectral phase state, and this can be applied to both Bell inequality violations[20,22] as well as quantum state tomography[22] of high-dimensional frequency entangled qudit systems[22] (Fig. 4). Both experiments, either 1) using unbalanced interferometers or 2) employing direct projection measurements with electro-optic modulation, demonstrate that energy-time entanglement covers the full comb. However, the electro-optic modulation approach provides direct access to the quantum modes and their superposition states. Building on this, different arrangements of the control elements (phase modulators, pulse shapers) can enable further quantum processing in the frequency-domain, discussed in the next section.

By combining the time- and frequency-bin concepts[53], it is feasible to generate hyper-entangled states that consist of at least four independent parties (qubits or qudits) with only two photons[52]. This becomes possible by using distinct time-scales for the frequency- and time-bin entanglements, which enables access to both encoding forms simultaneously.

Multi-photon entanglement[8], high-dimensional states[15] and hyper-entangled[92] states have all predominantly been demonstrated using large-scale free-space optics. Microcavity-based quantum frequency combs, on the other hand, can realize these states via integrated chips and/or fibers which offer enhanced scalability for large-scale quantum states.

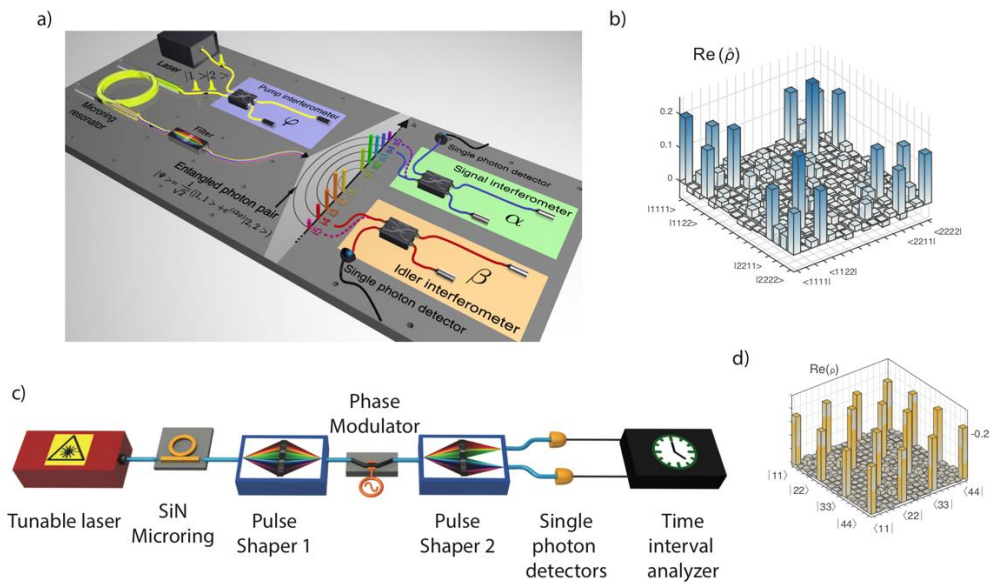

**Figure 4 | Complex quantum states from micro-ring quantum combs** a) Setup for time-bin multi-photon entangled states generation in frequency combs. b) Measured density matrix of a four-photon state. (adapted from Ref. 21) c) Experimental setup for the demonstration of frequency-bin entangled high-dimensional states (adapted from Ref. 20) d). Measured density matrix of a four-dimensional state (adapted from Ref. 22).

**Quantum processing with entangled frequency combs**

In order to realize the enormous potential of the quantum frequency combs detailed in the previous section, quantum gates that operate on the frequency degree of freedom are essential. In fact, the very advantages which make quantum frequency combs so attractive as scalable sources – namely their existence in a single spatial mode combined with their immunity to environmental perturbations – require new concepts in controlling them. Whereas path-encoded qubits in the spatial domain can be manipulated with beamsplitters and phase shifters[93,94], frequency-bin encoding demands energy-modifying operations for state manipulation. This is where classical devices from telecommunications, particularly Fourier-transform pulse shapers and electro-optic phase modulators, can play a key role, and indeed they have already had a significant impact on the field of quantum frequency combs[20,22,52,53,79,95–97].

Classical pulse shaping relies on the spectral decomposition of an input light field, followed by the application of a user-defined phase pattern in the generated Fourier plane, either with a fixed mask or liquid crystal pixels. When the optical frequencies of an input pulse are recombined, the result is an arbitrarily shaped temporal field[91]. However, pulse shapers may also be viewed more generally as user-programmable arbitrary spectral phase and amplitude filters. This provides the capability for manipulating the correlation properties of classical incoherent light fields[98] and, importantly, of quantum light. The time correlation function of entangled photons was first successfully shaped in 2005[83], with subsequent experiments extending to amplitude shaping[84], state characterization[97], and high-dimensional information encoding[85].

Electro-optic phase modulation represents the Fourier counterpart of pulse shaping – it applies a temporal, rather than spectral, phase pattern to the input[99]. Importantly, electro-optic modulation is independent of optical power and is linear in the optical field amplitude, and so it can be efficiently used to manipulate quantum light. Indeed, electro-optic modulation was first demonstrated on single photons in 2008[100], and subsequently applied to continuous energy-time correlations in nonlocal modulation cancellation[101], frequency-bin entanglement[79], and spread spectral encoding[102]. Recent experiments have also realized frequency shifting[103] and time lensing of single photons[104]. Although these focus on continuous photon spectra, or wavepackets – rather than discrete frequencies – they highlight the importance of electro-optics in single-photon control.

Pulse shaping can be straightforwardly applied to quantum frequency combs – a pulse shaper with sufficient resolution can impart an arbitrary phase shift to each comb-line, serving as a bank of phase shifters over all frequency bins. However, inducing interference between frequency bins (analogous to a spatial beamsplitter) is more challenging. As noted in the previous section, electro-optic phase modulators, when driven by an RF voltage commensurate with the frequency-bin spacing (FSR of the resonator), produce sidebands that allow neighboring bins to overlap (analogous to a multi-mode beamsplitter). However, due to the symmetric nature of electro-optic modulation driven by a single tone, any operation designed to interfere adjacent bins also yields unwanted sidebands outside of the modes of interest. Such 'spectral scattering' fundamentally limits the success probability of single-modulator-based operations, as is the case in the entanglement-verification schemes described in the previous section[20,22,95] as well as probabilistic frequency-domain Hong-Ou-Mandel interference[105]: they permit projections onto various frequency superpositions, but because of the intrinsic scattering loss, they are not unitary gates in the sense of transforming qubits within a common input/output computational space.

A solution to this involves cascading phase modulator/pulse shaper pairs. Initially proposed in the context of spectral linear-optical quantum computation[106], the basic idea is to synthesize frequency-bin unitary transformations via a series of temporal and spectral phase operations. These sequences generate probability amplitudes that are initially scattered outside of the encoding space and then caused to return into the encoding space before exiting the gate. This approach was shown to be sufficient to construct a universal gate set[107] consisting of a qubit phase gate, a qubit Hadamard (*H*) transformation, and a two-qubit controlled-Z gate (CZ) [106]. Scaling arguments imply that any unitary transformation can be achieved in this fashion, with the number of elements increasing only linearly with the dimension of the computational space[106]. In this way, the 'quantum frequency processor' (QFP)[96] forms an alternative unitary decomposition to the more conventional beamsplitter/phase-shifter approach in spatial encoding[93,94]. Figure 5 provides a schematic of how this approach could be applied in a gate for two frequency bins. Although the initial phase modulator spreads the probability amplitudes over bins outside of the two-dimensional input Hilbert space, the subsequent pulse shaper and modulator can reunite the scattered amplitudes via optical interference, yielding a closed single-qubit gate. The effective number of frequency bins populated in the middle of the operation depends on the specific manipulation, but for those examined so far has been on the order of 2-4 times the size of the computational space[106,108].

QFP gates have recently been demonstrated with the three-element configuration of Fig. 5 designed to synthesize a Hadamard gate (2x2 beamsplitter) over two frequency bins. Measured spectra for a 25 GHz-spaced QFP are shown in Fig. 6a[108]. Note that all combinations of bins 0 and 1 produce output states that are also in the qubit computational space, apart from small residual scattering (<3%) due to the use of a pure sinewave phase modulation rather than arbitrary RF patterns. The measured fidelity is $0.99998 \pm 0.00003$ confirming the high precision of this approach. Moreover, as noted above, the response of ideal phase modulators yields both upper and lower RF sidebands, independent of the center frequency of the input. While this is a challenge in terms of scattering photons into other frequency bins, it turns out to be ideal for parallelization. Specifically, this QFP was found to enable parallel Hadamard gates across the full (40 nm) optical C-band, valuable for accessing the broadband entanglement intrinsic to quantum frequency combs. Extending this approach to higher-dimensional operations is possible as well by, for example, using two-tone RF drive signals, realizing a frequency-bin tritter (3x3 beamsplitter or frequency mixer) in the same QFP[108].

The potential for parallel operations extends beyond the replication of the *same* gate over many bins – it is also possible to realize *different* gates on multiple qubits in the same spatial mode. For example, by simply adjusting the pulse shaper phase, the reflectivity of the frequency beamsplitter described above can be tuned smoothly between 50% (which yields the Hadamard gate, *H*) and 0% (which yields the identity operation, I). This allows the QFP to operate jointly on spectrally separated qubits with distinct single-qubit gates. Figure 6b provides an example with entangled frequency-bin qubits[96]. When only one qubit experiences the *H* gate, the initial spectral correlations are eliminated. Yet, when both undergo an *H* rotation, correlations are recovered, but with the opposite sign. This suggests the significant potential for performing massively parallelized and arbitrary state rotations on quantum frequency combs. As with any optical approach to quantum information processing, two-photon *entangling* operations are extremely difficult, and yet the probabilistic designs developed in spatial encoding[94] can be successfully applied to frequency bins, as exemplified in a recent coincidence-basis controlled-NOT (CNOT)[109] gate. Alternatively, by exploiting hyper-entanglement in energy and time, single-photon two-qudit gates have also been realized[52,53]. Since the hyper-entangled gates are deterministic, they provide, in conjunction with other linear-optical quantum computation approaches, a powerful tool for achieving high-dimensional quantum information processing with entangled frequency combs[52,53].

A key challenge in the QFP paradigm is the bandwidth required of the electro-optic modulators used for phase modulation. Phase modulators have reached ~100 GHz[110] with commercial devices being typically limited to 50 GHz. Although using high-order sidebands can extend this bandwidth, such modulation schemes suffer from low efficiency, thus severely limiting the circuit *depth* (the number of operations which can be concatenated sequentially) as well as qubit (or qudit) *connectivity* (the maximum separation of frequency bins that can be efficiently mixed by a single modulator). While nonlinear optical processes can support wider bandwidth operation[75,111–114], a more efficient approach is moving to smaller FSR combs[20]. Classical micro-combs with FSR spacings below 25 GHz have been reported[115–117] – enabling the coupling of several frequency modes using the fundamental RF tone of a single electro-optic modulator.

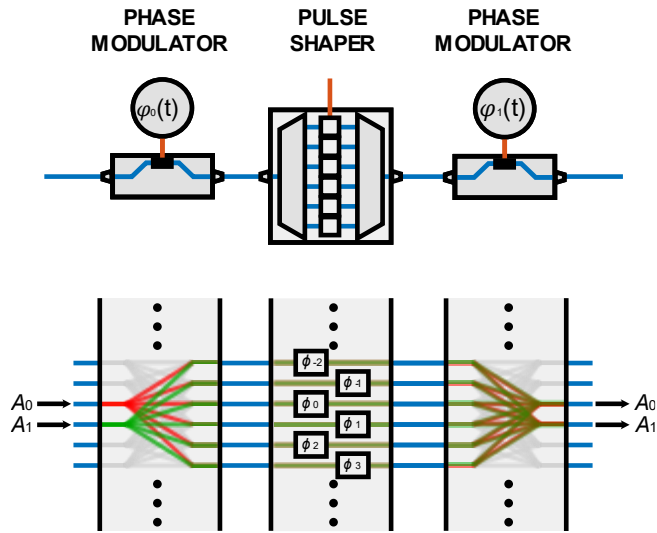

**Figure 5 | Concept of quantum frequency processor (QFP) with three elements.** The physical configuration (top) comprises phase modulators and a central pulse shaper. The modulators are driven by phase patterns periodic at the frequency-bin spacing, and the pulse shaper applies arbitrary phase shifts to each bin. In the conceptual picture (bottom), frequency bins (rails) are mixed by the modulators and receive specific phase shifts by the pulse shaper. In this example, the cascade ensures probability amplitudes from both modes $A_0$ (red) and $A_1$ (green) return to the qubit space on exiting, after experiencing the desired operation.

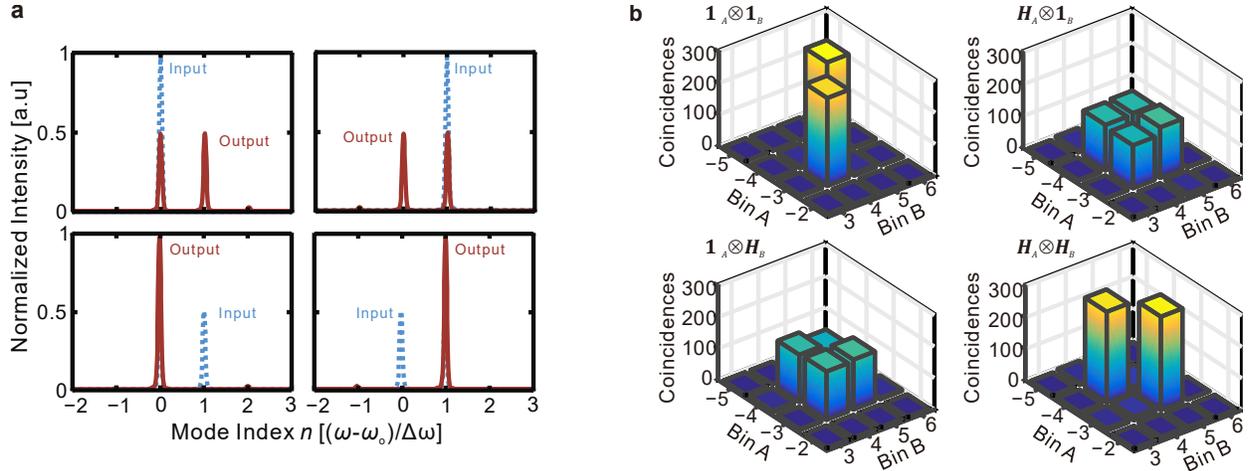

**Figure 6 | Experimental examples of frequency-bin gates. a**, Output spectra of a Hadamard gate for the inputs $|0\rangle$ (upper left), $|1\rangle$ (upper right), $|0\rangle+|1\rangle$ (lower left), and $|0\rangle-|1\rangle$ (lower right). **b**, measured correlations of a two-qubit frequency-bin entangled state, where each qubit is operated on by either the identity or Hadamard independently. Figure reproduced with permission from: **a**, Ref. [108] [© APS 2018]; **b**, Ref. [96] [© OSA 2018].

**Summary and outlook**

Quantum science requires increasingly more complex and large-scale quantum resources, for testing fundamental novel quantum physics as well as realizing relevant and meaningful non-classical signal processing tasks. Solid state and atom-based quantum systems (e.g. trapped ions, superconducting qubits) have been shown to be highly controllable and capable of achieving efficient multi qubit interactions[11,118,119]. Considerable effort has been devoted to scaling up their information content to large quantum systems, but this has proven challenging. Photonic systems, on the other hand, while not particularly suited for deterministic two qubit interactions because of the weak interaction between photons[120], can provide many quantum modes simultaneously in e.g. polarization, path, and time-frequency. Furthermore, photons are useful for control, transport and communication in other quantum platforms[121,122], and in fact they are currently the only option for some applications (such as quantum networks). The intrinsic spectral multimode property introduced by quantum frequency combs provides scalability that can compensate in part for the lack of photon interaction efficiency and is thus of significant interest.

The quantum frequency combs reviewed here offer a powerful and versatile platform to generate complex, multi-mode states in a scalable manner. They provide a unique framework for the manipulation of quantum states in a single spatial mode using standard fiber optic telecommunication components. While current micro-comb based quantum systems are typically limited to relatively few (~40) frequency modes, and single-photon manipulation schemes to mode counts of 3-4, they are scalable by employing smaller FSR resonators and using more advanced low-loss integrated electro-optic modulation[110] and integrated spectral phase manipulation schemes[123,124]. Further research and development in broadband detection techniques for frequency-encoded photon states[83,125] will further advance their use. Quantum photonic integrated circuits will naturally benefit from on-going advancements in classical telecommunications integrated circuits and components as well as enhanced access to custom designs with the potential for reduced loss through integrated photonics foundry services.

As for all sources of quantum photonic states based on nonlinear spontaneous processes, micro-combs are fundamentally stochastic in nature. Efficient quantum photonic systems will benefit from approaches that create effective deterministic sources[81], where quantum frequency combs allow novel approaches that exploit their multimoded nature[75]. However, in contrast to previous sources, quantum frequency combs provide a large quantum resource per photon which could bring the potential to compensate for the drawbacks of non-determinism[52]. Apart from that, the challenges for creating deterministic quantum frequency comb systems are very similar to those of deterministic single photon sources.

Quantum frequency combs have achieved significant breakthroughs in the generation and control of highly complex photon states that will be critical for large-scale quantum information processing. It is clear that, whichever roles are ultimately assumed by photonic-based quantum information processing, whether it be for interlinks, communications, simulations or computing, those systems will tremendously benefit from being able to generate and operate on large-scale multi-mode quantum states, where micro-combs have indeed demonstrated an attractive and powerful approach towards achieving this fundamental goal.

**Acknowledgments**
We thank J. Azaña, L. Caspani, P. Roztocki, S. Sciara, Y. Zhang, N. Lingaraju and P. Lougovski for discussions. We acknowledge funding from: Canada Research Chairs (MESI PSR-SIIR); Natural Sciences and Engineering Research Council of Canada (NSERC); H2020 Marie SkłodowskaCurie Actions (MSCA) (656607); ITMO Fellowship and Professorship Program (08-08); 1000 Talents Sichuan Program; Australian Research Council (ARC) (DP150104327); John Templeton Foundation (JTF) number 60478; U.S. Department of Energy, Office of Science, Office of Advanced Scientific Computing Research Quantum Algorithm Teams; Laboratory Directed Research and Development Program of Oak Ridge National Laboratory, managed by UT-Battelle, LLC, for the U.S. Department of Energy; National Science Foundation under award number 1839191-ECCS.